\documentclass[letterpaper,12pt]{article}
\usepackage{tabularx} 
\usepackage{amsmath}  
\usepackage[margin=1in,letterpaper]{geometry} 
\usepackage{cite} 
\usepackage[final]{hyperref} 
\usepackage{authblk}
\usepackage{colortbl}
\usepackage{url}
\usepackage[pdftex]{graphicx}
\hypersetup{
	colorlinks=true,       
	linkcolor=blue,        
	citecolor=blue,        
	filecolor=magenta,     
	urlcolor=blue         
}

\newcommand{\Tabref}[1]{Table\,\ref{#1}}
\newcommand{\Figref}[1]{Fig.\,\ref{#1}}
\newcommand{\Equref}[1]{Eq.\,\ref{#1}}

\begin{document}

\title{Performance evaluation of a GEM-based readout module for the ILC TPC with a large aperture GEM-like gating device by a beam test}
\author{Yumi Aoki
\\
on behalf of the LCTPC Asia group}
\affil{SOKENDAI}

\maketitle

\begin{abstract}
  
   A high momentum resolution is required for the precision  measurement of Higgs boson at the International Linear Collider (ILC) using the recoil mass technique.  The International Large Detector (ILD) is designed to meet this requirement by an MPGD-readout Time Projection Chamber (TPC) providing about 200 sample points each with a spatial resolution of 100\,$\mu$m operated in a magnetic field of 3.5\,T. However, there is a potential problem that many  positive ions generated in the gas amplification process in the end-plane detector modules would flow back into the drift volume of the TPC and distort its electric field. These positive ions must be removed by a gating device before reaching the drift volume. 
    We have developed a GEM-like gating device (gating foil) to prevent ions from back-flowing to the drift volume and evaluated its performance. The  performance measurement was carried out at DESY, using  a 5\,GeV electron beam and the Large Prototype TPC in a 1~T magnet field. We have measured the spatial resolution of our MPGD module equipped with the gating foil and the electron transmission rate of the gating device. This was the world first test beam experiment of a ``wireless'' TPC equipped with a high performance gating device.
   In this report, we present our results on the spatial resolution  and the electron transmission rate. \footnote{Talk presented at the International Workshop on Future Linear Colliders (LCWS2019), Sendai, Japan, 28 October-1 November, 2019. C19-10-28.}

\end{abstract}

\section{Introduction}
The International Linear Collider (ILC)~\cite{Aus1}  is an electron-positron linear collider with its first stage being a 250\,GeV Higgs factory to study the properties of Higgs bosons to an unprecedented precision. The International Large Detector (ILD) is one of the two proposed detectors for the ILC. This study discusses the performance evaluation of the Time Projection Chamber (TPC), which is the central tracker of the ILD. 

This report is organized as follows. 
In section 2, we set the performance goal for the ILD TPC. Section 3 introduces the ion feedback problem and proposes our solution, a GEM-like gating device. Section 4 describes our test beam experiment at DESY and section 5 gives results of our test beam data analysis. Finally section 6 summarizes out main results and concludes this report.

\section{ILD TPC and Its Performance Goal}
The centeral tracker of the ILD is an MPGD-readout TPC designed to reconstruct tracks of densely packed jets of particles and determine their momenta and particle spices through dE/dx information. The ILD TPC has a radius of 1.8\,m and drift length of 2.35\,m~\cite{Aus2} on each side and is filled with the T2K gas consisting of 95\% Argon, 3\% CF$_4$, and 2\% iso-butane. The ILD TPC is operated in a solenoidal magnetic field of 3.5\,T. Figure 1 shows an artist's view of the structure of the ILD TPC.

\begin{figure}[ht] 
        \centering \includegraphics[width=0.8\columnwidth]{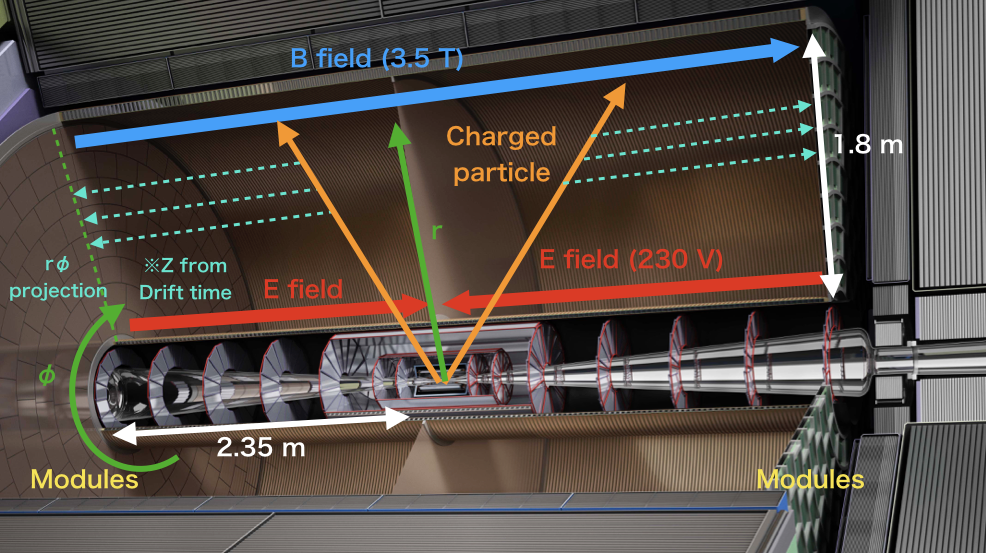}
        \caption{
                \label{fig:1} 
                Tracking in TPC.  
        }
\end{figure}

According to the baseline design described in \cite{Aus1}, the momentum resolution of the ILD TPC is required to be better than\\
\begin{eqnarray}
\frac{\Delta p}{ p^{2}} = 1 \times 10^{-4}(\mathrm{GeV} / \mathrm{c})^{-1}. 
\label{Equ:1}
\end{eqnarray}

We can translate this performance requirement on the momentum resolution to TPC design parameters using the Gl\"{u}ckstern formula ~\cite{Aus2} : \\
\begin{eqnarray}
\frac{\sigma_{P_{T}}}{P_{T}}\simeq \sqrt{\left(\frac{\alpha'\sigma_{x}}{BL^2}\right)^2 \left(\frac{720}{n+4}\right)P_{T}^2 + \left(\frac{\alpha'C}{BL}\right)^2 \left(\frac{10}{7}\left(\frac{X}{X_{0}} \right)\right)},
\label{Equ:10}
\end{eqnarray}
where $B$ is the magnetic field, $L$ is the track length, $n$ is the number of sampling points, $\sigma_x$ is the spatial resolution per hit point, ${\alpha}^\prime$ is $\frac{1}{c}$, C is the constant (13.6 GeV), and $\frac{X}{X_{0}}$ is thickness measured in radiation length units. When we substitute ILC TPC parameters:$n=220$, $B=3.5\,T$, and $L=$1.5 m in this formula, we get our spatial resolution goal: $\sigma_{r \phi}<100~\mu$m.\\
\label{Equ:2}
\section{Ion Feedback Problem}
\subsection{Ion Feedback Problem and Gating Foil}
\begin{figure}[ht] 
        \centering \includegraphics[width=0.6\columnwidth]{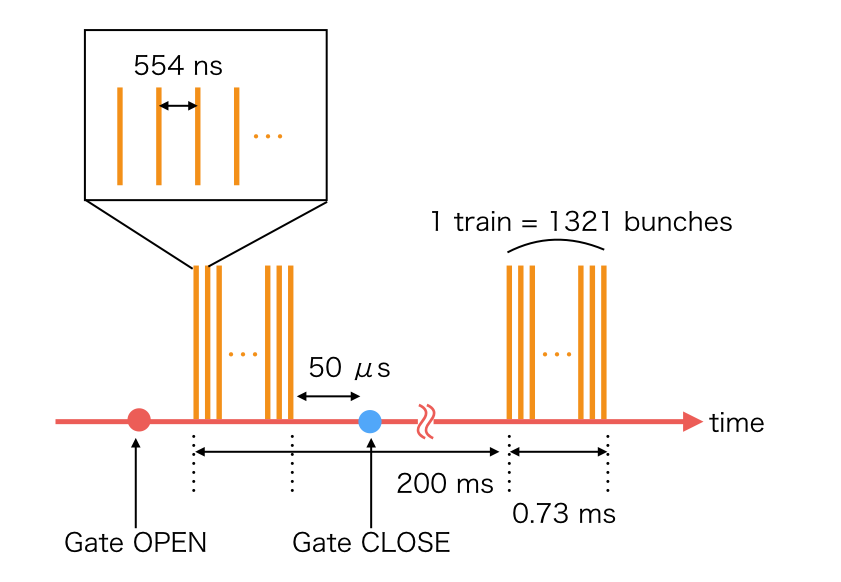}
        \caption{
                \label{fig:2} 
               The bunch structure of ILC 
        }
\end{figure}

Positive ions created by gas amplification would back-flow into the drift volume and distort the electric field and consequently drift paths of track electrons, thereby deteriorating spatial resolution, unless some gating device is used. The distribution of the positive ions is determined by the bunch structure of the accelerator. Figure\,\ref{fig:2} shows the ILC bunch structure.  

The ions for a single bunch train form a disk with about 1\,cm thickness. Since the ion drift velocity is O(1000) times slower than that of electrons, there will be up to 3 ion disks in the drift volume. The hit point distortion due to the 3 ion disks has been estimated to be greater than 60\,$\mu$m~\cite{Aus3}, which is comparable to the required spatial resolution and hence necessitates a gating device. 

\subsection{A Large Aperture GEM-like Gating Device}
To solve the ion feedback problem, we have developed a large aperture GEM-like gating device (the gating foil) with FUJIKURA company\cite{Ref:FUJIKURA}. 
\begin{figure}[ht] 
       \centering \includegraphics[width=0.4\columnwidth]{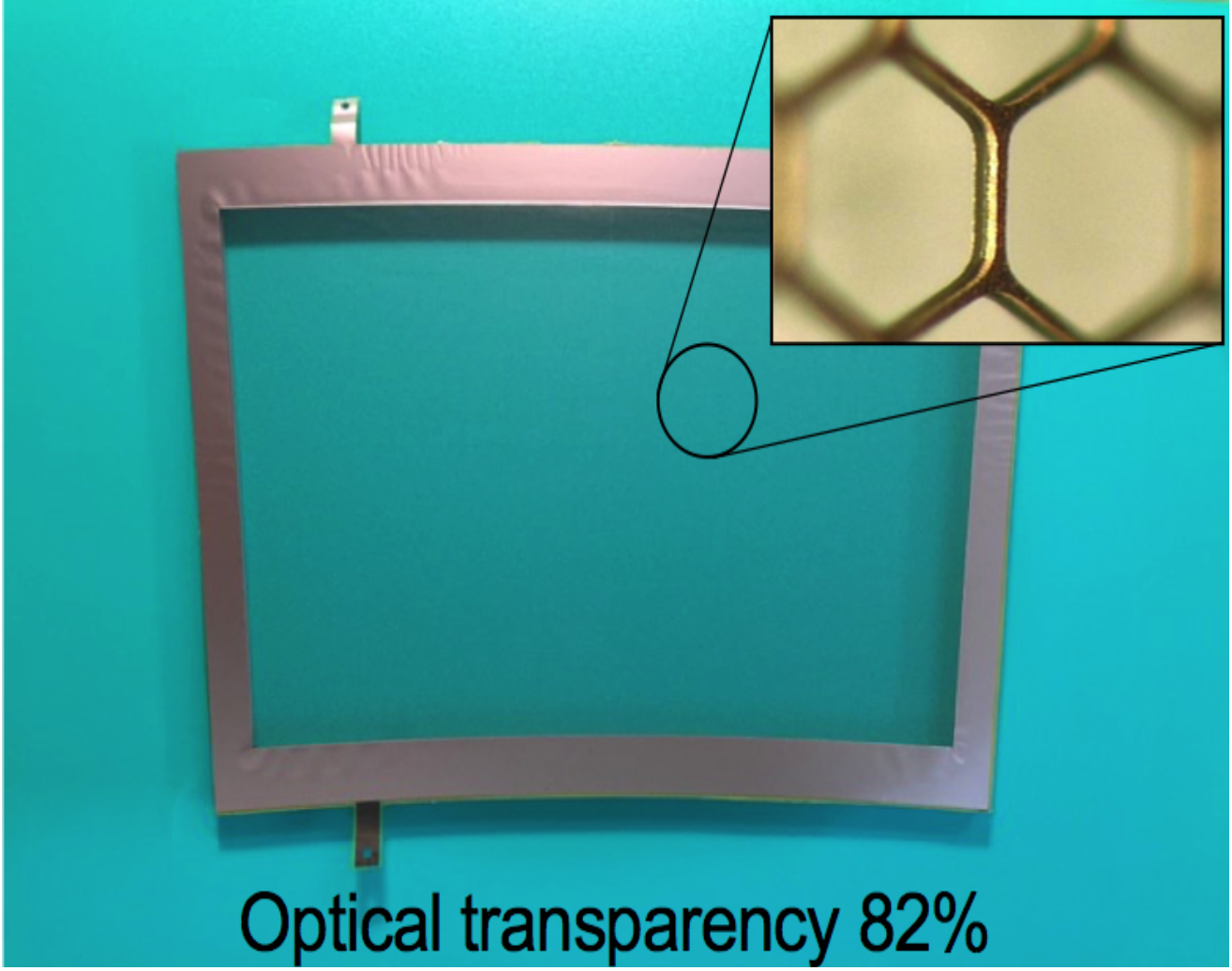} \includegraphics[width=0.4\columnwidth]{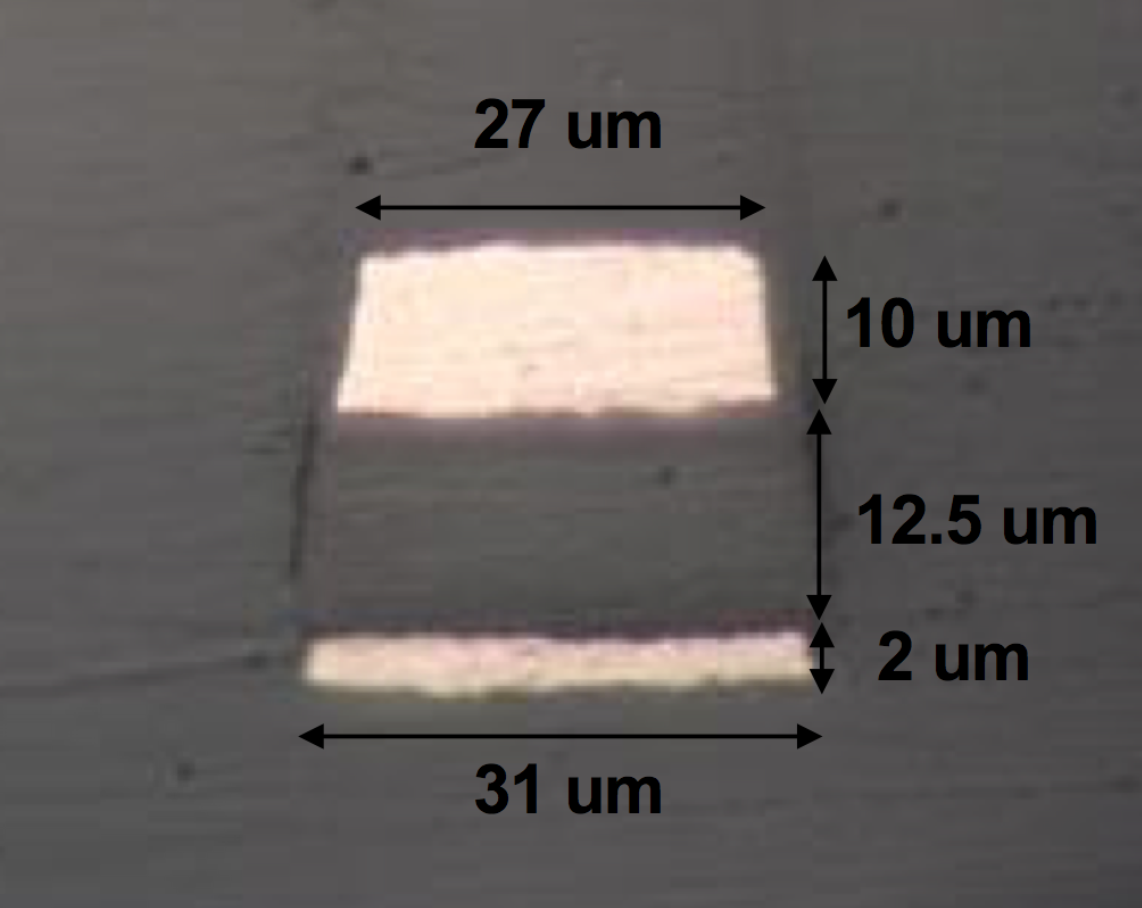}

        \caption{
                \label{fig:3} 
               The gating foil. 
        }
\end{figure}
It is like a Gas Electron Multiplier (GEM), but is very thin (24.5\,$\mu$m) having a honeycomb structure shape to realize bigger holes so as to achieve a high electron transmission rate. The optical transparency is about 82\,$\%$.  Before a bunch train comes, we open the gating foil by applying the voltage on the copper electrodes in the normal direction.  After the bunch train goes through, we close the gate by reverting the voltage to absorb the positive ions.


\section{Beam Test}
We carried out a beam test at DESY, using the large prototype of TPC to confirm the performance of our GEM module with the gating foil. This is the first beam test using the module with the gating foil.

\subsection{Set-up}
The test beam experiment was carried out at DESY T24 beam line~\cite{ref:desy}. We used a large prototype TPC (LP1) installed in a 1\,T superconducting solenoidal magnet called PCMAG from KEK~\cite{ref:lp1}. LP1 has a diameter of 77\,cm and a maximum drift length of 56.8\,cm filled with T2K gas. It has an end plate that can house up to seven readout modules. In our experiment we mounted two GEM-readout modules, one with and the other without a gating foil, fabricated by the Asian LCTPC group.

\begin{figure}[ht] 
        \centering \includegraphics[width=0.7\columnwidth]{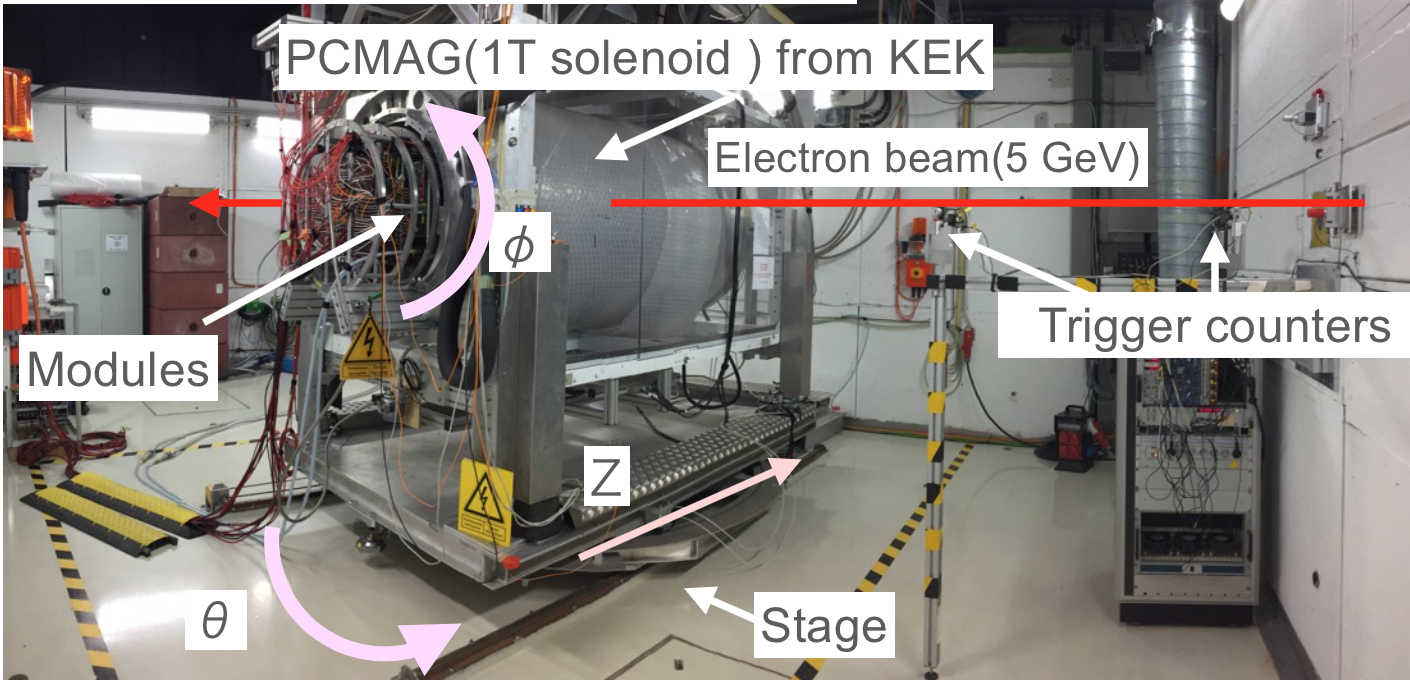}
        \caption{
                \label{fig:4} 
               DESY test facility
        }
\end{figure}

The 5\,GeV electron beam passes through two pairs of trigger counters before entering LP1. PCMAG is mounted on a movable stage so that we can change drift distance ($z$) and two angles, $\theta$ and $\phi$ as shown in Fig.4. The Asian GEM modules has a double stack of 100\,$\mu$m thick amplification GEMs with an anode plane of about 17\,cm $\times$ 20\,cm, which is segmented into 28 rows consisting of readout pads of about 1\,mm $\times$ 5.6\,mm. The module without the gating foil has a field shaper instead so as to keep the electric field uniformity near the module.
The signal from each anode pad was readout using the ALTRO readout system\cite{Ref:ALTRO} and recorded by a DAQ PC.
The drift field during the experiment was $3.55$\,V/cm.

\begin{figure}[ht] 
        \centering \includegraphics[width=0.6\columnwidth]{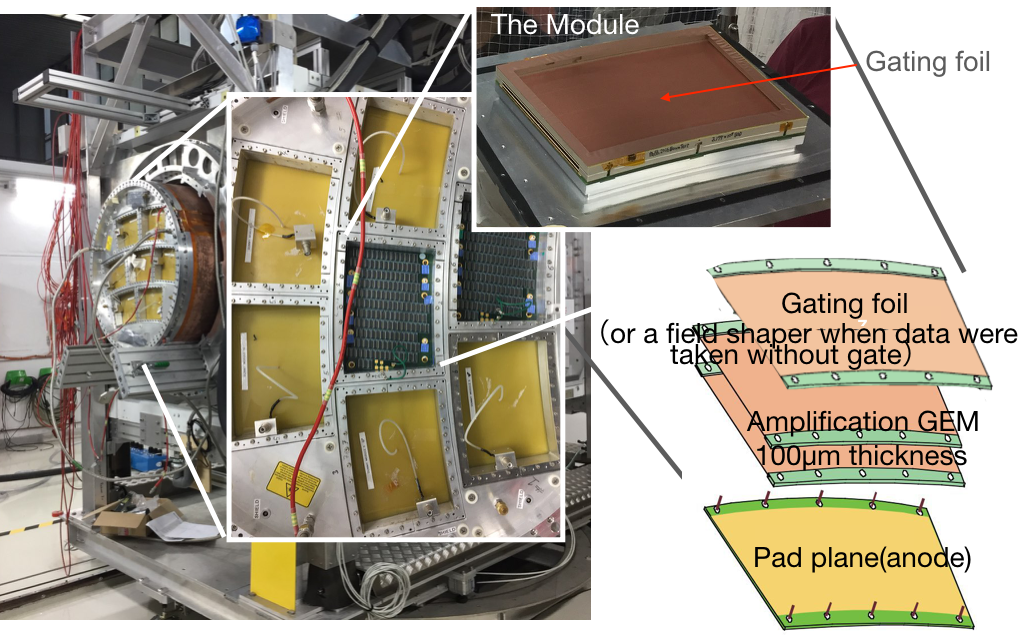}
        \caption{
                \label{fig:5} 
               The module setup
        }
\end{figure}

\clearpage
\subsection{Data Set Used in This Analysis}
We took data at different drift lengths, different angles and different gating foil voltages with and without the gating foil. Applied voltages on various electrodes of the GEM module are graphically shown in Fig.\,6. We use the data shown in ~\Tabref{tbl:1} in this study, where $V_{gate}$ is the voltage applied to the gating foil  at which the electron transmission is expected to be maximal.

\begin{table}[htbp]
\begin{center}
\caption{The data set}
\label{tbl:1} 
\begin{tabular}{|c||c|c|c|} 
\hline
 Center module & With gating foil &  Without gating foil\\
\hline
\end{tabular}

\begin{tabular}{|c||c|c|} 
\hline
 Z[cm](drift length) & 2.5, 5, 7.5, 10, 12.5, 15, 20, 25, 30, 35, 40, 45, 50, 55\\
$\phi$[degree] & 0 \\
$\theta$[degree] & 0 \\
$V_{gate}$[V] & 3.5 \\
B(magnetic field)[T] & 1 \\
Beam & 5 GeV electron beam\\
Gas &T2K gas (Ar : $\rm CF_4$ : Iso-$\rm C_4 \rm H_{10}$ = 95 : 3 : 2 [$\%$])\\
Framework & MarlinTPC (20000event/1 run)\\
\hline
\end{tabular}

\end{center}
\end{table}

\begin{figure}[ht] 
        \centering \includegraphics[width=0.8\columnwidth]{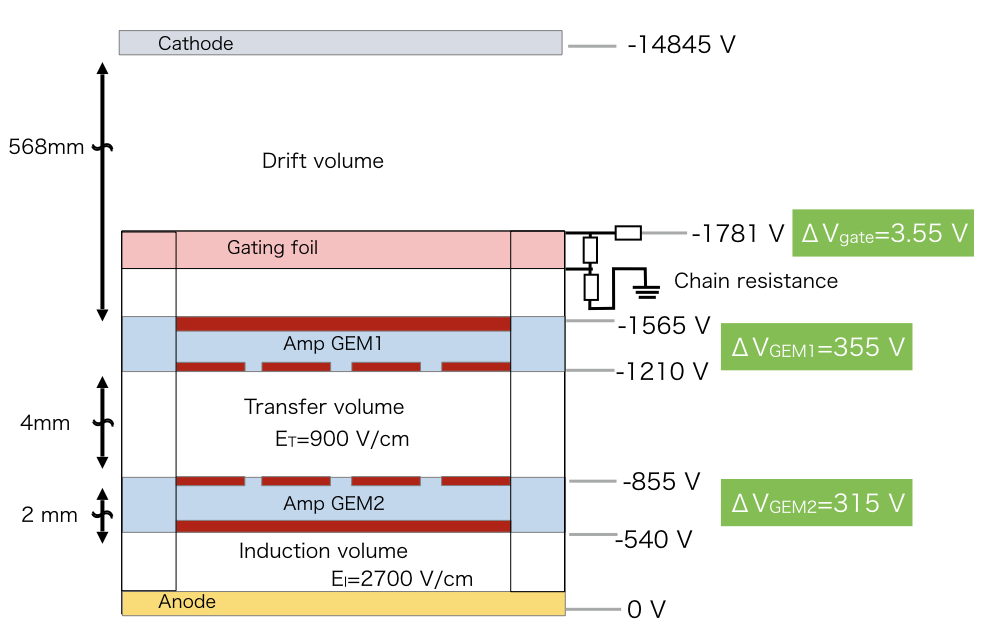}
        \caption{
                \label{fig:6} 
               The voltage setup
        }
\end{figure}

\begin{figure}[ht] 
        \centering \includegraphics[width=0.7\columnwidth]{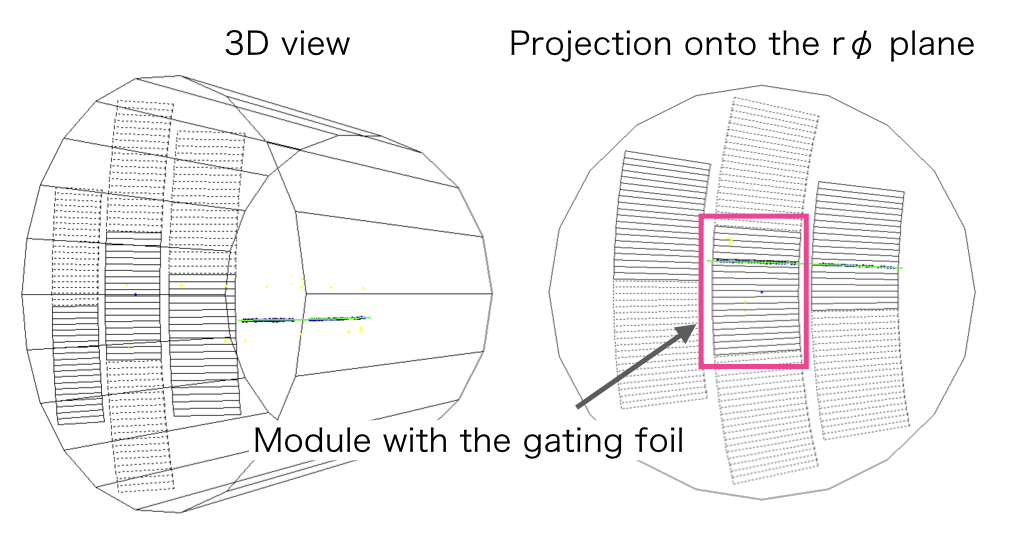}
        \caption{
                \label{fig:7} 
               A typical event as seen by our GEM module with the gating foil. 
        }
\end{figure}
Throughout the data taking, the beam was adjusted, as seen in Fig.\,7, to go through our GEM module in the region
far enough from the module boundaries to avoid potential edge effects due to possible electric field distortions.

\section{Data Analysis and Results}
\subsection{Event Selection}
We required the angle between the track projected onto the pad plane and the line perpendicular to the pad row in question to be less than 0.03\,rad to exclude inclined tracks in order to avoid angular pad effects. We also eliminated events with multiple tracks caused by electromagnetic showers created upstream. See Fig.8 for the distributions of the track angle and the number of tracks per event. 

\begin{figure}[ht] 
        \includegraphics[width=0.5\columnwidth]{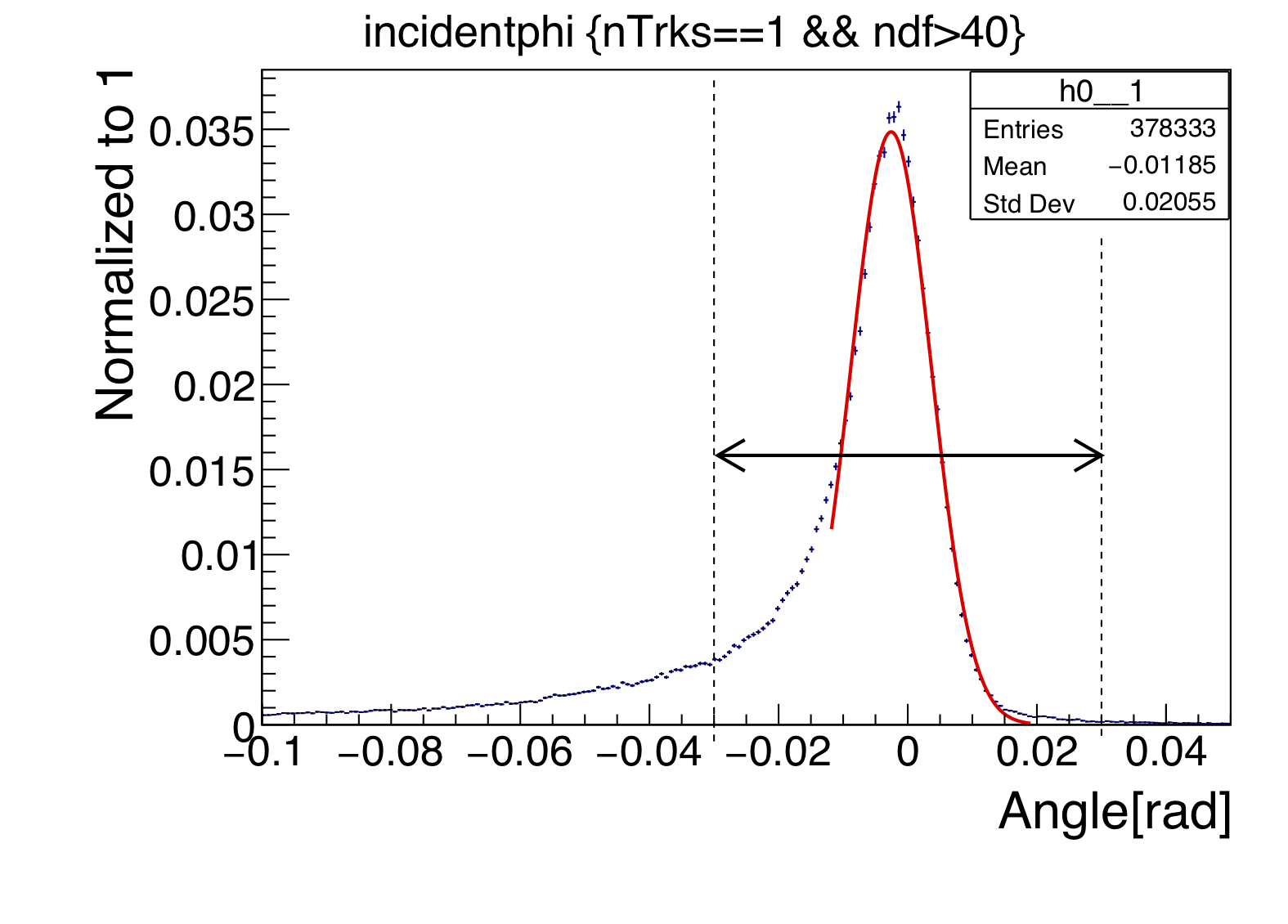}
        \includegraphics[width=0.5\columnwidth]{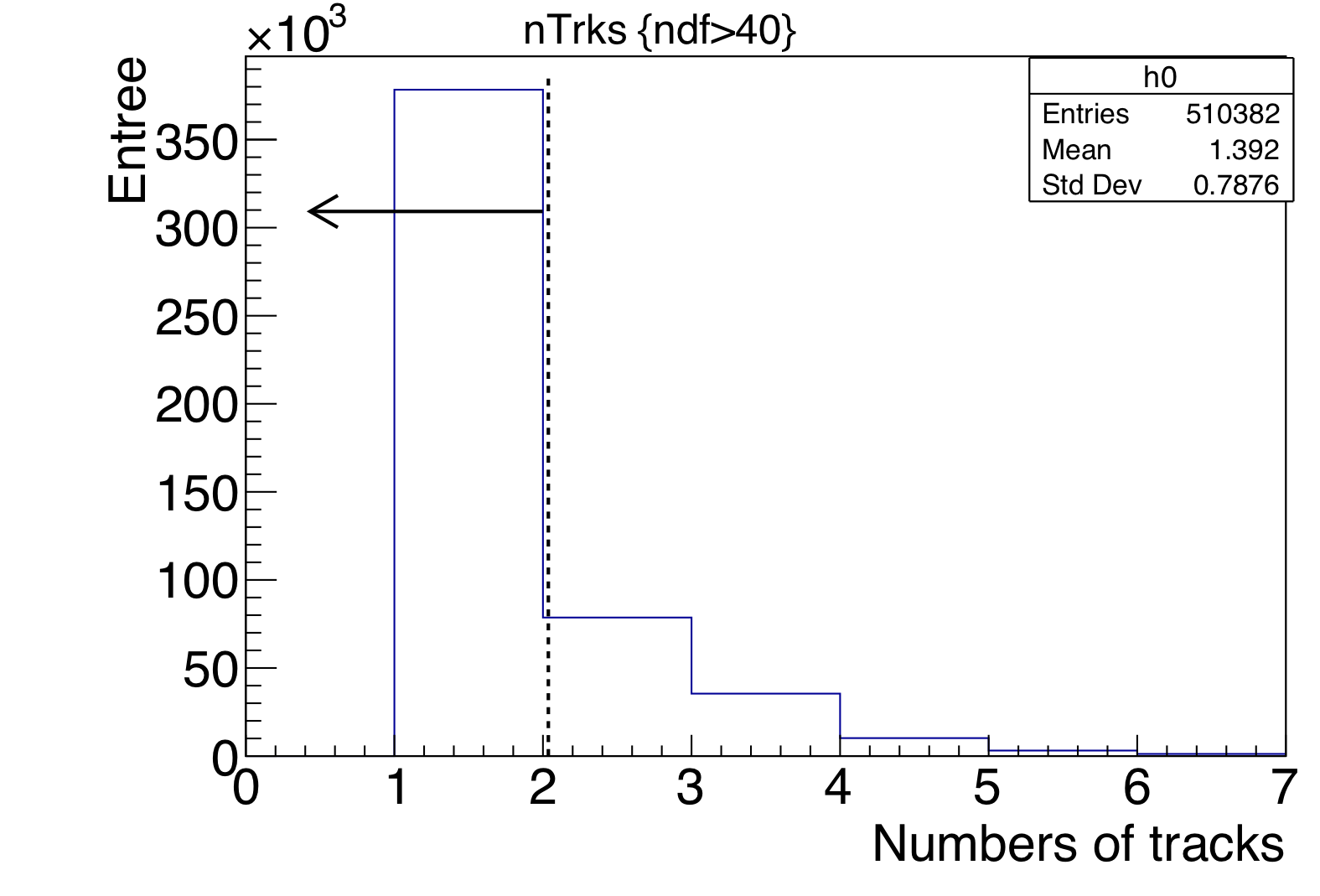}
        \caption{
                \label{fig:8} 
               Event selection cuts}
\end{figure}
\clearpage
\subsection{Single Hit Spatial Resolution}
Our primary purpose of the test beam experiment was to make sure that the gating foil is sufficiently transparent for track electrons so that the spatial resolution stays high enough with the gating foil.

\begin{figure}[ht] 
        \centering \includegraphics[width=0.8\columnwidth]{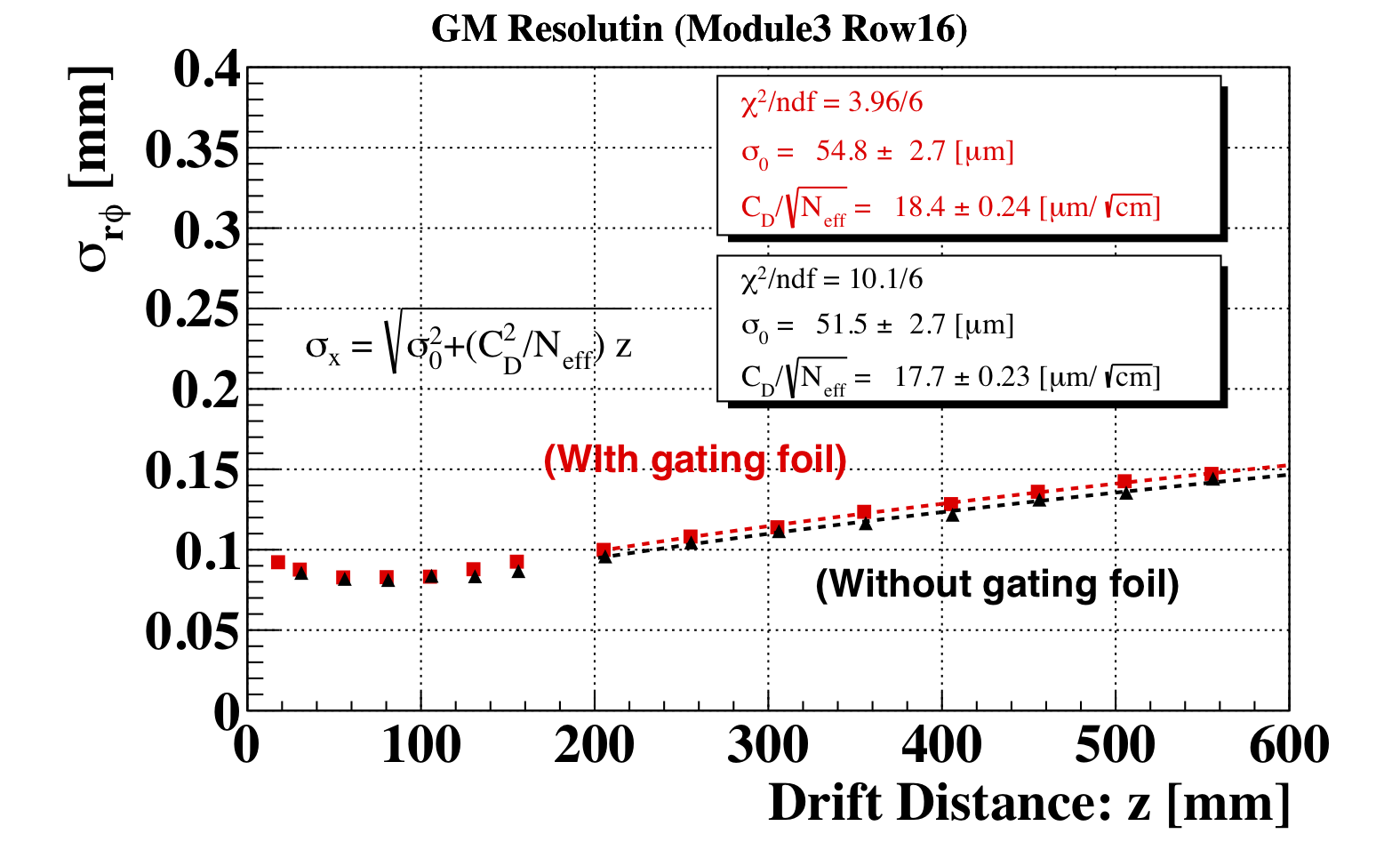}
        \caption{
                \label{fig:11} 
               Single hit spatial resolution as a function of drift distance with (red) and without (black) the gating foil.}
\end{figure}

In order to estimate the single hit spatial resolution, we first plotted the residual between the position of the hit in question and the track with and without the hit included in track fitting. Each of the resultant two residual distributions was then fitted to a gaussian to get its standard deviation, $\sigma_{r \phi (in)}$ or $\sigma_{r \phi (out)}$, depending on whether the hit in question was included in the fit or not. Finally we calculated the single hit resolution as $\sigma_{r \phi}=\sqrt{\sigma_{r \phi(in)} \sigma_{r \phi(out)}}$ and plotted it as a function of the drift distance as shown in \Figref{fig:11}. As seen in this figure, the difference was found to be less than 10\,\%, small enough to insure the required performance. 

\subsection{Pad Response}
\begin{figure}[ht] 
        \centering \includegraphics[width=0.8\columnwidth]{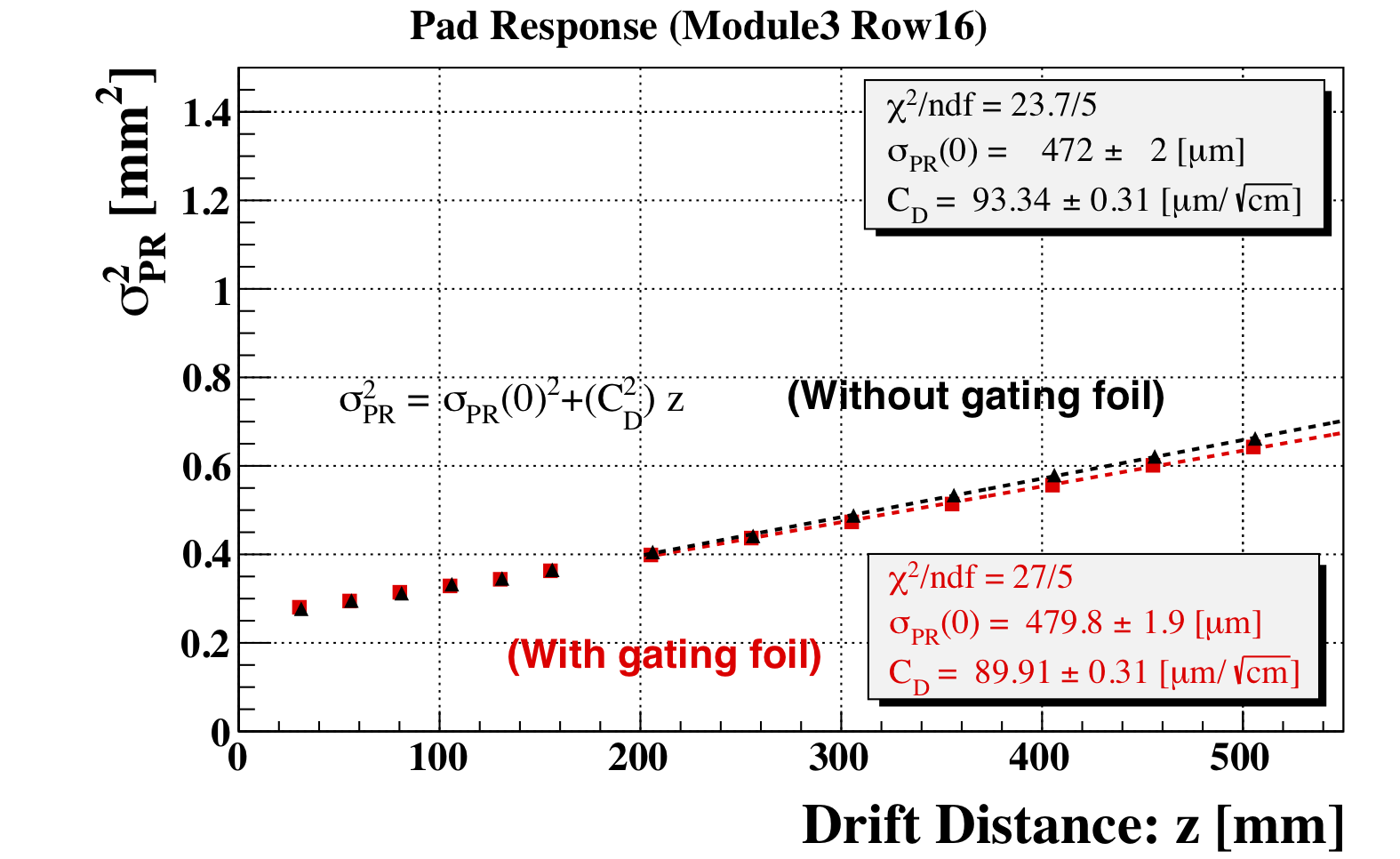}
        \caption{
                \label{fig:10} 
               Width of the pad response function plotted against drift distance with (red) and without (black) the gating foil.
        }
\end{figure}

We evaluated the diffusion constant from the width of the pad response function plotted as a function of drift distance $z$. The width (standard deviation) of the pad response function was calculated by fitting the distribution of the charge fraction on $i$-th pad: $Q_i / \sum_i Q_i$ against $r \phi_{\text {hit }}-r \phi_{\mathrm{i}\text {-th pad center }}$, where $r$ is the radius of the pad row in question, $\phi_{\rm hit}$ and $\phi_{\mathrm{i}\text {-th pad center }}$ are the azimuthal angles (about the center of the pad row arc) of the hit and the $i$-th pad center, respectively. 
The obtained width of the pad response function is plotted as a function of drift distance in \Figref{fig:10} for data taken with (red) and without (black) the gating foil.

We get the diffusion constant by fitting the following formula to this result: 
\begin{eqnarray}
\sigma_{P R}^{2}=\sigma_{P R}(0)^{2}+\left(C_D^{2}\right) z \\
\sigma_{P R}(0) = \sigma_{\mathrm{PRF}}^{2}+\frac{w^{2}}{12},
\label{Equ:4}
\end{eqnarray}
where $C_D$ is the diffusion constant, $\sigma_{\mathrm{PRF}}$ is the intrinsic charge signal width on the pad plane, and $w$ is the pad pitch. 
The resultant $C_D$ values with (red line) and without (black line) the gating foil are 89.91~$\mu m / \sqrt{cm}$ and 93.34~$\mu \rm m / \sqrt{cm}$, respectively. The difference of these two values seems significant. 
For the same drift field and the same gas conditions (i.e. mixing ratio, pressure, and temperature), the $C_D$ values are expected to be same. In order to evaluate how much the temperature and pressure could have affected the diffusion constant, we calculated potential  $C_D$ value change by garfield++~\cite{Ref:garfield} simulation for the temperature and pressure differences between the two data sets. The simulation result is shown in \Tabref{tbl:10}. 

The $C_D$ difference cannot be explained by the gas condition. A remaining possible cause of the $C_D$ difference is beam intensity dependence of positive ions effects. 
\begin{table}[htbp]
\begin{center}
\caption{The result of garfield++ simulation for the temperature and pressure differences between the two data sets. }
\label{tbl:10} 
\begin{tabular}{|c|c|c|}\hline & {\text { with gating foil }} & {\text { without gating foil }} \\ \hline 
\text { Temperature[K] } & {291.28} & {290.4} \\ \hline 
\text { Pressure[hPa]} & {1010.79} & {1005.31} \\ \hline 
{$C_D$}  {[$\mu$m/$\sqrt{cm}$]} & {94.0 $\pm$ 0.2} & {94.2 $\pm$ 0.3} \\ \hline
\end{tabular}
\end{center}
\end{table}

\clearpage
\subsection{Estimation of Electron Transmission}
In order to quantitatively estimate the electron transmission of our gating foil, we now compare the effective numbers of track electrons ($N_{\rm eff}$) that are used for hit position measurements with and without the gating foil. The long drift distance behavior of the single hit spatial resolution is given by
\begin{eqnarray}
\sigma_{r \phi}=\sqrt{\sigma_{0}^{2}+\frac{\left(C_D^{2}\right)}{N_{\rm eff}} z},
\label{Equ:5}
\end{eqnarray}
where $\sigma_0$ is a constant depending on the pad geometry and the pad response function, and $N_{\rm  eff}$ is the effective number of electrons determined by ionization statistics and gas gain fluctuation and expressed by
 \begin{eqnarray}
N_{\rm eff}=\left[\left\langle\frac{1}{N}\right\rangle\left\langle\left(\frac{G}{\bar{G}}\right)^{2}\right\rangle\right]^{-1}. 
\label{Equ:11}
\end{eqnarray}
In this expression, $N$ is the number of the track electrons per pad row, $G$ is the gas gain, and ${\bar{G}}$ is its average. 
Fitting \Equref{Equ:5} to the single hit spatial resolutions plotted against drift distance and using the $C_D$ values obtained in the previous subsection, we can now extract $N_{\rm eff}$. \Tabref{tbl:2} summarizes $C_D$, $C_D / \sqrt{N_{\rm eff}}$, and resultant ${N_{\rm eff}}$ values.
\begin{table}[htbp]
\begin{center}
\caption{ $C_D$ and $C_D / \sqrt{N_{\rm eff}}$, $N_{\rm eff}$ value and the ratio of  $N_{\rm eff}$}
\label{tbl:2} 
\begin{tabular}{|c|c|c|c|}
\hline & {\text { w/gate }} & {\text { w/o gate }} & {\text { ratio[\%] }} \\ \hline 
\text { $C_D$ [$\mu m /\sqrt{cm}$]} & {89.91 $\pm$ 0.37} & {93.34+0.31}  &\\ \hline
 \text { $C_D / \sqrt{N_{\rm eff}}$ [$\mu m /\sqrt{cm}$] } & {18.6 $\pm$ 0.3} & {17.8 $\pm$ 0.3} &\\ \hline
  \text { $N_{\rm eff}$ } & {23.9 $\pm$ 0.7} & {27.8 $\pm$ 0.8} & {85.9 $\pm$ 3.3} \\ \hline
\end{tabular}
\end{center}
\end{table}

We can then estimate the electron transmission rate by taking the ratio of $N_{\rm eff}$ values obtained with and without the gating foil: 
\begin{eqnarray}
R_{e.t.} \approx \frac{N_{\rm eff}(\mathrm{w} / \mathrm{Gate})}{N_{\rm eff}(\mathrm{w} / \mathrm{o}~\mathrm{Gate})}.  
\label{Equ:7}
\end{eqnarray}

We found the electron transmission rate to be 85.9$\pm 3.3\,\%$, which is consistent with the optical transparency of the gating foil of about 82 $\%$.  

\section{Summary and Conclusion}
    We have developed a GEM-like gating device (gating foil) to prevent positive ions from back-flowing into the ILD TPC drift volume.  Its performance was measured at DESY, using  a 5\,GeV electron beam and the large prototype TPC in a 1\,T magnet field. We have measured the spatial resolution of our MPGD module equipped with the gating foil and estimated the electron transmission rate of the gating foil to be 85.9$\pm 3.3\,\%$, high enough to insure the spatial resolution required for the ILD TPC. 
       
\section*{Acknowledgements}
This study was supported by the LCTPC collaboration. The production of the gating GEM is cooperative effort by Fujikura Ltd. The author would like to thank the members of the LCTPC group and Fujikura Ltd. In particular, P.\,Colas， S.\,Ganjour, R.\,Diener, O.\,Sch\"afer, F.\,M\"uller, L.\,J\"onsson, U.\,Mj\"ornmark, and D.\,Arai deserve special mention. Special thanks also go to Y. Makida, M. Kawai, K. Kasami and O. Araoka of the KEK IPNS cryogenic group, and A. Yamamoto of the KEK cryogenic centre for their support in the configuration and installation of the superconducting PCMAG solenoid. The authors would like to thank the technical team at the DESY II accelerator and test beam facility for the smooth operation of the test beam and the support during the test beam campaign. The contributions to the experiment by the University of Lund, KEK, Nikhef and CEA are gratefully acknowledged.

\end{document}